\documentclass[12pt]{article}
\usepackage{a4wide,epsfig,amsmath,amssymb,cite,scalefnt}
\usepackage[dvips]{feynmp}

\newcommand{\gsim}{\;\rlap{\lower 3.5 pt \hbox{$\mathchar \sim$}} \raise 1pt
 \hbox {$>$}\;}
\newcommand{\lsim}{\;\rlap{\lower 3.5 pt \hbox{$\mathchar \sim$}} \raise 1pt
 \hbox {$<$}\;}

\newcommand{\LH}{L_H}
\newcommand{\Lt}{L_t}
\newcommand{\taums}{\bar{\tau}}

\begin{document}

\title{\vskip-3cm{\baselineskip14pt
    \begin{flushleft}
      \normalsize SFB/CPP-07-46\\
      \normalsize TTP07-20
%  \\
%      \normalsize arXiv:YYMM.NNNN
  \end{flushleft}}
  \vskip1.5cm
  Higgs Decay to Gluons at NNLO
}
\author{\small Marco Schreck and Matthias Steinhauser\\[1em]
  {\small\it Institut f{\"u}r Theoretische Teilchenphysik,
    Universit{\"a}t Karlsruhe (TH)}\\
  {\small\it 76128 Karlsruhe, Germany}}

\date{}

\maketitle

\thispagestyle{empty}

\begin{abstract}
  We present an analytical calculation of the 
  next-to-next-to-leading order corrections
  to the partial decay width $H\to gg$ for a Higgs boson in the intermediate
  mass range. We apply an asymptotic expansion for 
  $M_H\ll 2M_t$ and compute three terms in the expansion. 
  The leading term confirms the results present in the literature. It
  is argued that our result is equivalent to an exact calculation up
  to $M_H\approx M_t$. For a Higgs boson mass of 120~GeV the
  power-suppressed terms lead to corrections of about 9\%
  in the next-to-next-to-leading order coefficient.

  \medskip

  \noindent
  PACS numbers: 14.80.Bn  12.38.-t

\end{abstract}

\thispagestyle{empty}

\newpage

%- }}} Header:
%- {{{ Introduction:

\section{\label{sec::intro}Introduction}

Up to date the Standard Model (SM) Higgs boson has evaded its direct
detection. Electroweak precision data collected at the CERN Large
Electron-Positron Collider (LEP), at the Stanford Linear Collider (SLC)
and at the Tevatron at Fermilab predict a light SM
Higgs boson below approximately 200~\mbox{GeV}. On the other hand, the
direct search at LEP2 has excluded Higgs bosons below 114~GeV which
leaves a relatively narrow window for the mass. 
A Higgs boson in this mass range, often also referred to as
intermediate-mass Higgs boson, predominantly decays into bottom quarks
and $W$ bosons depending whether the latter decay is kinematically
allowed. 
A further decay channel which is of phenomenological interest is the
one into gluons. At lowest order this process is mediated by a top
quark loop~\cite{Ellis:1975ap,Shifman:1979eb}. 
The next-to-leading order (NLO) corrections are quite
large and amount to roughly 
70\%~\cite{Inami:1982xt,Dawson:1991au,Djouadi:1991tk,Graudenz:1992pv,Dawson:1993qf,Spira:1995rr}.
About ten years ago the NNLO corrections have been computed in the
limit of an infinitely heavy top quark
mass~\cite{Chetyrkin:1997iv,Steinhauser:2002rq}. Since the 
corrections amount to approximately 20\% they increased the
reliability of the perturbative expansion of the decay rate. This has been
further substantiated by the recent evaluation of the NNNLO
corrections~\cite{Baikov:2006ch}, again in the infinite top quark mass
limit, which provides a contribution of about 2\%.

The calculations to NNLO and NNNLO are both based on an effective theory
which results from the SM after integrating out the top quark. The
effective Lagrangian is constructed from operators with dimension five
which is sufficient to obtain the leading term in the heavy-top
expansion.
The inclusion of higher-dimensional operators would lead to
power-suppressed terms. However, the renormalization of the corresponding 
effective Lagrangian is quite tedious. In this paper we follow a
different approach and consider the Higgs-boson propagator in the full
SM. Its imaginary part leads --- with the help of the optical theorem
--- directly to the decay rate.
The disadvantage of this method is that at NNLO five-loop
corrections to the Higgs two-point function have to be considered
since the LO result already requires a three-loop diagram
where the coupling of the Higgs boson to the gluons is mediated by 
two top quark triangles. They are connected by two gluon lines.
The advantage of our method relies on the straightforward
automatization using state-of-the-art techniques and program packages
as we will show in Section~\ref{sec::tecdet}.

An important feature of the Higgs boson decay into gluons is
its affinity to the production mechanism in the gluon fusion channel.
For the latter the NNLO corrections are again only known within the
framework of the effective
theory~\cite{Harlander:2002wh,Anastasiou:2002yz,Ravindran:2003um}. 
The calculation performed in the present paper gives a first hint
about the potential
size of the terms which are suppressed by the top quark mass.

Let us finally mention that the corresponding NNLO corrections to the
decay of the Higgs boson into two photons have been obtained in
Ref.~\cite{Steinhauser:1996wy}. Note that in this case a simple
expansion of the Higgs-photon-photon vertex diagrams is sufficient to obtain 
power-suppressed terms the heavy $M_t$-limit. Thus only three- instead
of five-loop diagrams have to be considered.

The paper is organized as follows: In Section~\ref{sec::tecdet} we
present some details of our calculation and Section~\ref{sec::res}
contains our analytical results and the discussion about the numerical
implications. We conclude the paper with Section~\ref{sec::concl}.

%- }}}
%- {{{ The calculation:

\section{\label{sec::tecdet}The calculation}

\begin{figure}[t]
  \begin{center}
    \begin{tabular}{c}
      \includegraphics[width=0.3\textwidth]{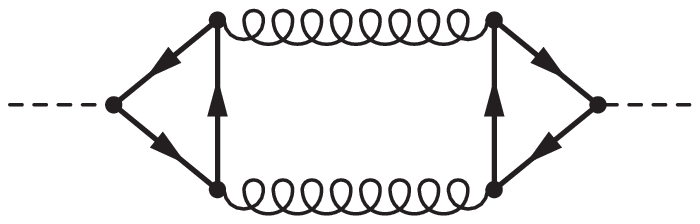}
      \mbox{}\hfill\mbox{}
      \includegraphics[width=0.3\textwidth]{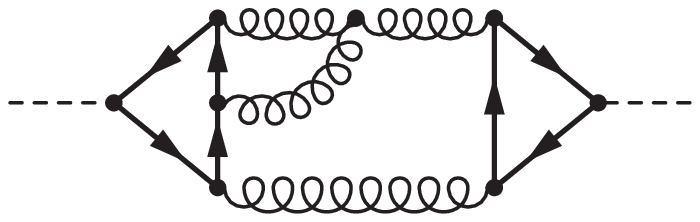}
      \mbox{}\hfill\mbox{}
      \includegraphics[width=0.3\textwidth]{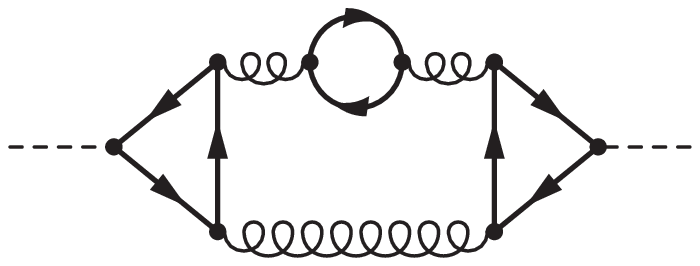}
      \\
      \mbox{}\hfill(a)\hspace*{11em}(b)\hspace*{11em}(c)\hfill\mbox{}
      \\
      \includegraphics[width=0.3\textwidth]{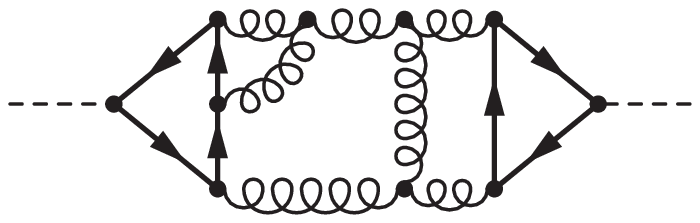}
      \mbox{}\hspace*{2em}\mbox{}
      \raisebox{-.45em}{
      \includegraphics[width=0.3\textwidth]{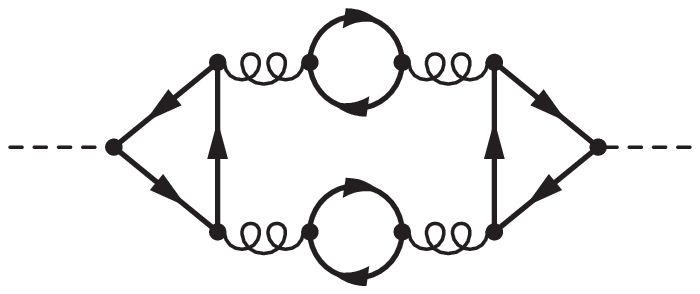} 
      }
      \\
      \mbox{}\hfill(d)\hspace*{14em}(e)\hfill\mbox{}
    \end{tabular}
    \caption{\label{fig::diags}
      Sample diagrams contributing to $\Pi_H$. The curly lines
      represent gluons and the solid lines stand for light quarks or
      top quarks. 
      The Higgs boson (external dashed lines) only couples to top quarks.
    }
  \end{center}
\end{figure}

As already mentioned in the Introduction, for the computation of the
decay rate $\Gamma(H\to gg)$ we consider the Higgs boson self energy,
$\Pi_{H}(q^2)$,  and apply the optical
theorem which in our case reads
\begin{eqnarray}
  \Gamma(H\to gg) &=& \frac{1}{M_H} \mbox{Im}\left[ \Pi_{H}(M_H^2+i0) \right]
  \,.
\end{eqnarray}
Since we do not consider the (exclusive) decay into light quarks
only those diagrams contribute to $\Pi_{H}$ where the 
Higgs boson couples to the top quark. Thus the LO
result is obtained from the three-loop diagram shown in
Fig.~\ref{fig::diags}(a) (and the one with crossed gluons).
The NLO and NNLO results are obtained by dressing this diagram with
additional gluons and (light and heavy) quark loops. Sample diagrams
can be found in Fig.~\ref{fig::diags}(b)--(e).
It is clear that beyond LO next to gluons also light quarks can be
produced. In particular, at NNLO also a final state
$q\bar{q}q^\prime\bar{q}^\prime$ with light quark flavours $q$ and
$q^\prime$ is possible.

\begin{figure}[t]
  \begin{center}
%    \begin{tabular}{ccc}
%      \includegraphics[width=0.3\textwidth]{figs/Dia04.eps} 
%      & \raisebox{1.9em}{$\Longrightarrow$}
%      & \includegraphics[width=0.5\textwidth]{figs/Exp01.eps} \\
%      & \raisebox{2.0em}{+}
%      & \includegraphics[width=0.5\textwidth]{figs/Exp02.eps} \\
%      & \raisebox{2.1em}{+}
%      & \includegraphics[width=0.5\textwidth]{figs/Exp03.eps} \\
%      & \raisebox{2.2em}{+}
%      & \includegraphics[width=0.5\textwidth]{figs/Exp04.eps} \\
%      & \raisebox{2.3em}{+}
%      & \includegraphics[width=0.5\textwidth]{figs/Exp05.eps}
%    \end{tabular}
    \leavevmode
    \epsfxsize=\textwidth
    \epsffile[80 270 510 580]{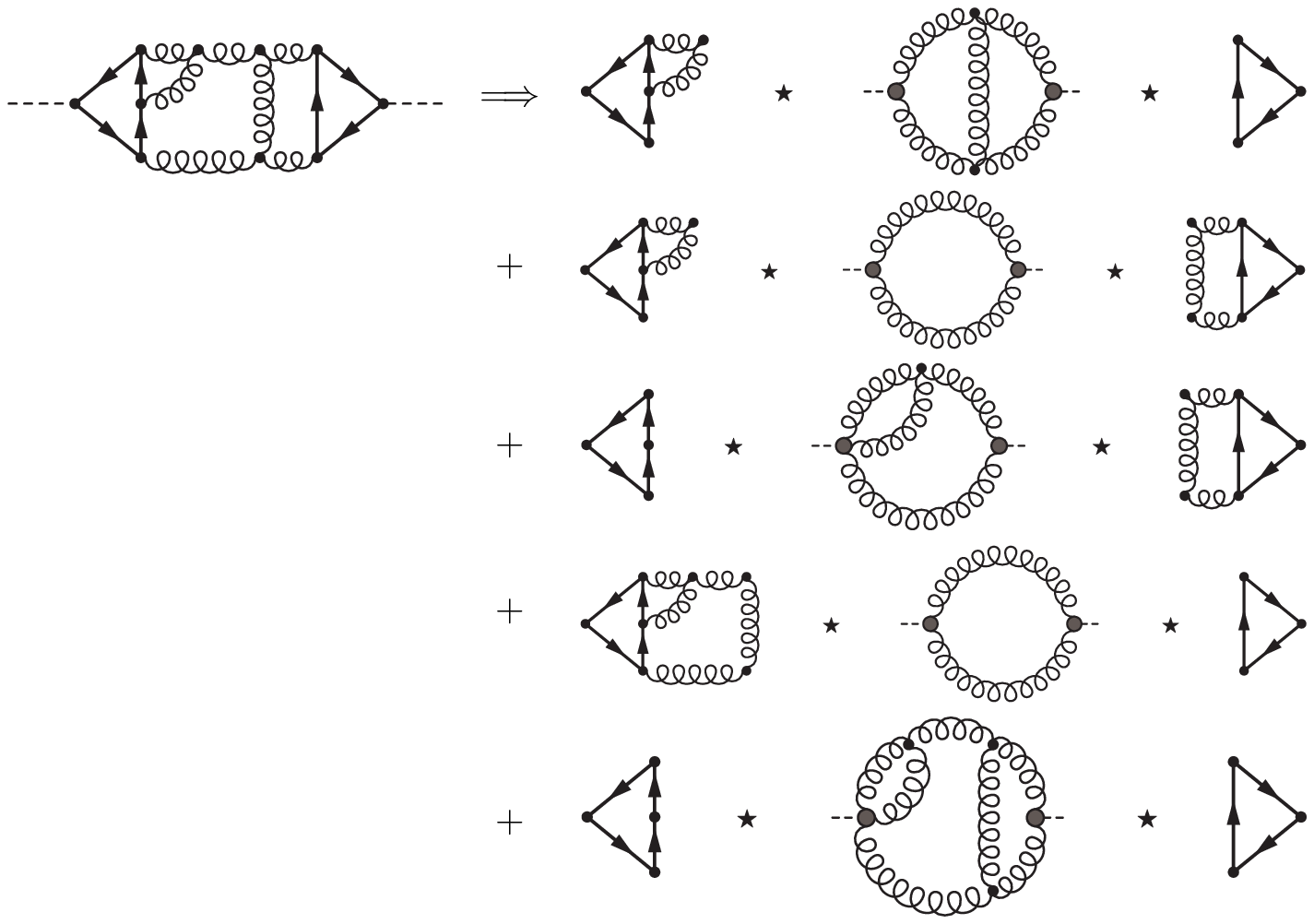}
    \caption{\label{fig::hmp}
      Sample five-loop diagram contribution to $\Pi_H$
      where the hard-mass procedure is applied in 
      graphical form. Only the subgraphs contributing to the imaginary
      part are shown.
    }
  \end{center}
\end{figure}

Our approach for the computation of $\Gamma(H\to gg)$ requires the
evaluation of five-loop diagrams in order to obtain the NNLO
corrections. With the currently available techniques their exact
evaluation is out of reach. However, for a Higgs boson in the
intermediate mass range it is promising to consider an expansion
for $M_H^2\ll M_t^2$ which shows a rapid convergence at LO and NLO
as we demonstrate in Section~\ref{sec::res}.
In this limit we can apply the so-called hard-mass procedure 
(see, e.g., Ref.~\cite{Smirnov:2001}) to the propagator diagrams 
which then factorize into products of one-, two- and
three-loop integrals.
A graphical example is shown in Fig.~\ref{fig::hmp} where ---
according to the rules of the hard-mass procedure --- five subdiagrams
are identified. They have to be expanded in the small quantities and
then to be inserted in the co-subgraph which (in this example)
only consists of gluons.
In Fig.~\ref{fig::hmp} they are sandwiched between the symbols ``$\star$''.
The hard-mass procedure has been automated in the program {\tt
  exp}~\cite{Harlander:1997zb,Seidensticker:1999bb} which we use for
our calculation. 

At LO we have two contributing Feynman diagrams, at NLO there are
already 71 and at NNLO 2649.
All Feynman diagrams are generated with {\tt
QGRAF}~\cite{Nogueira:1991ex} and the various topologies are identified
with the help of {\tt q2e}~\cite{Harlander:1997zb,Seidensticker:1999bb}. 
Afterwards {\tt exp}~\cite{Harlander:1997zb,Seidensticker:1999bb} applies
the hard-mass expansion and produces output which can 
be read by the {\tt FORM}~\cite{Vermaseren:2000nd} packages {\tt
  MATAD}~\cite{Steinhauser:2000ry} and {\tt
  MINCER}~\cite{Larin:1991fz}. {\tt MATAD} can compute vacuum 
integrals up to 
three-loop order where in our case the mass scale is given by the top
quark. On the other hand, {\tt MINCER} can handle massless
propagator-type diagrams where for our application the external
momentum is given by the Higgs boson mass.

According to the optical theorem we have to compute the imaginary part
of $\Pi_H$ which requires the evaluation of the pole part of the
three-, four- and five-loop diagrams. Since {\tt MATAD} and {\tt
  MINCER} are constructed such that at most the finite parts of
three-loop diagrams can be computed slight modifications are necessary
which allow the correct evaluation of the $1/\epsilon$
pole part of the five-loop diagrams.

For some of the diagrams it turned out to be crucial to use the 
parallel versions of {\tt FORM} --- {\tt ParFORM}~\cite{Tentyukov:2006pr}
and {\tt TFORM}~\cite{Tentyukov:2007mu} --- in order to obtain results in
a moderate amount of time. With our setup, which is not tuned for a
parallel computation, we obtain speedup factors between four and five using
eight processors.

%- }}}
%- {{{ Results and discussion:

\section{\label{sec::res}Results and discussion}

%- {{{ LO and NLO:

\subsection{LO and NLO result}

The method described in the previous Section can be tested at LO and
NLO where the exact results are known. Let us for this purpose 
introduce the notation
\begin{eqnarray}
  \Gamma(H\to gg) &=& \Gamma_0 \left(\frac{\alpha_s(\mu)}{\pi}\right)^2
  \left[ \Gamma^{\rm LO}_{gg} 
  + \frac{\alpha_s(\mu)}{\pi} \Delta\Gamma^{\rm NLO}_{gg}
  + \left(\frac{\alpha_s(\mu)}{\pi}\right)^2
  \Delta\Gamma^{\rm NNLO}_{gg} + \ldots
  \right]
  \,,
\end{eqnarray}
with $\Gamma_0 = G_F M_H^3/(36\pi\sqrt{2})$.
$\Gamma^{\rm LO}_{gg}$ is given by~\cite{Ellis:1975ap,Shifman:1979eb}
\begin{eqnarray}
  \Gamma^{\rm LO}_{gg}
  &=&  \left[\frac{3}{2\tau}\left(1+\left(1-\frac{1}{\tau}\right)
    \arcsin^2\sqrt{\tau}\right)\right]^2 
  \nonumber\\
  &=& 
  1+\frac{7}{15}\tau+\frac{1543}{6300}\tau^2
  +\frac{226}{1575}\tau^3+\frac{55354}{606375}\tau^4
  +\frac{1461224}{23648625}\tau^5
  +\ldots
  \,,
  \label{eq::gamLO}
\end{eqnarray}
with $\tau=M_H^2/(4M_t^2)$ where $M_t$ is the on-shell quark mass. 
In the second line of
Eq.~(\ref{eq::gamLO}) we have performed an expansion up to order
$\tau^5$ which we could confirm by applying the hard-mass
procedure to the three-loop diagrams (cf. Fig~\ref{fig::diags}(a)).
The ellipses indicate higher order terms in $\tau$.

The exact NLO result is only known in numerical 
form~\cite{Dawson:1991au,Djouadi:1991tk,Graudenz:1992pv,Dawson:1993qf,Spira:1995rr}. 
Thus we
concentrate on the expansion for $M_H\ll2 M_t$. The first three
terms have been computed in Ref.~\cite{Larin:1995sq} which
we could confirm. Furthermore, we have added two more terms and arrive
at
\begin{eqnarray}
  \Delta\bar{\Gamma}^{\rm NLO}_{gg}
  &=& h_0^{\rm{nlo}}+h_1^{\rm{nlo}}\taums+h_2^{\rm{nlo}}\taums^2
  +h_3^{\rm{nlo}}\taums^3+h_4^{\rm{nlo}}\taums^4
  +h_5^{\rm{nlo}}\taums^5+\ldots
  \,,
  \label{eq::gamnlo}
\end{eqnarray}
where
\begin{eqnarray}
  h_0^{\rm{nlo}} &=&
  \frac{95}{4} + \frac{11}{2}\LH + 
  n_l \left(-\frac{7}{6} - \frac{\LH}{3}\right)
  \,,\nonumber\\
  h_1^{\rm{nlo}} &=&
  \frac{5803}{540} + 
  \frac{77}{30}\LH - \frac{14}{15}\Lt 
  + n_l\left(-\frac{29}{60} - \frac{7}{45}\LH\right)
  \,,\nonumber\\
  h_2^{\rm{nlo}} &=&
  \frac{1029839}{189000} + \frac{16973}{12600}\LH 
  - \frac{1543}{1575}\Lt 
  + n_l\left(-\frac{89533}{378000} 
  - \frac{1543}{18900}\LH\right)
  \,,\nonumber\\
  h_3^{\rm{nlo}} &=&
  \frac{9075763}{2976750} 
  + \frac{1243}{1575}\LH 
  - \frac{452}{525}\Lt
  + n_l\left(-\frac{3763}{28350} - \frac{226}{4725}\LH\right)
  \,,\nonumber\\
  h_4^{\rm{nlo}} &=&
  \frac{50854463}{27783000} 
  + \frac{27677}{55125}\LH 
  - \frac{442832}{606375}\Lt + 
   n_l \left(-\frac{10426231}{127338750} - 
   \frac{55354}{1819125}\LH\right)
  \,,\nonumber\\
  h_5^{\rm{nlo}} &=&
  \frac{252432553361}{218513295000} 
  + \frac{730612}{2149875}\LH 
  - \frac{2922448}{4729725}\Lt 
  \nonumber\\&&\mbox{}
  + n_l\left(-\frac{403722799}{7449316875} 
  - \frac{1461224}{70945875}\LH\right)
  \,,
  \label{eq::hggnlo}
\end{eqnarray}
with $L_H=\ln(\mu^2/M_H^2)$ and $L_t=\ln(\mu^2/m_t^2)$. In
Eq.~(\ref{eq::gamnlo}) we have $\bar{\tau}=M_H^2/(4 m_t^2)$
where $m_t\equiv m_t(\mu)$ is the $\overline{\rm MS}$ 
top quark mass.
For convenience we have kept the label referring to closed light-quark 
loops which takes the numerical value $n_l=5$.
Furthermore, we have $\alpha_s(\mu)\equiv\alpha_s^{(5)}(\mu)$,
i.e., the top quark has been decoupled from the running of the strong
coupling (see, e.g., Ref.~\cite{Steinhauser:2002rq}).

Transforming the top quark mass from the $\overline{\rm MS}$ to the
on-shell scheme~\cite{Gray:1990yh} leads to the following result which is evaluated in
numerical form 
\begin{eqnarray}
  \Delta\Gamma^{\rm NLO}_{gg}
  &=&
  17.9167 + 3.83333 L_H 
    + \left(9.574 + 1.789 L_H\right) \tau 
    \nonumber\\&&\mbox{}
    + \left(5.571 + 0.939 L_H\right) \tau^2  
    + \left(3.533 + 0.550 L_H\right) \tau^3
    \nonumber\\&&\mbox{}
    + \left(2.395 + 0.350 L_H\right) \tau^4  
    + \left(1.708 + 0.237 L_H\right) \tau^5
    + \ldots
  \,.
\end{eqnarray}

\begin{figure}[t]
  \begin{center}
    \begin{tabular}{c}
      \includegraphics[width=0.5\textwidth]{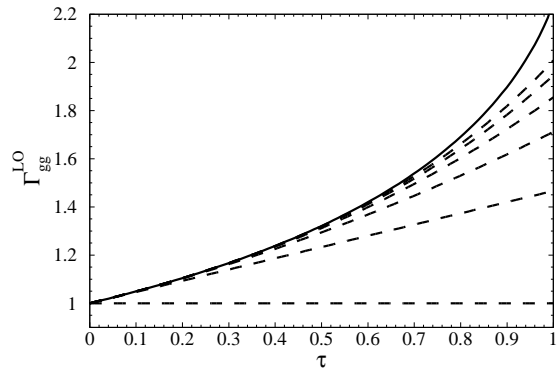}
    \end{tabular}
    \caption{\label{fig::lo}$\Gamma^{\rm LO}_{gg}$ as a function of
    $\tau$ where 
    successively higher orders are included (dashed lines).
    The exact result is shown as a solid line.
    }
  \end{center}
\end{figure}

\begin{figure}[t]
  \begin{center}
    \begin{tabular}{cc}
      \includegraphics[width=0.47\textwidth]{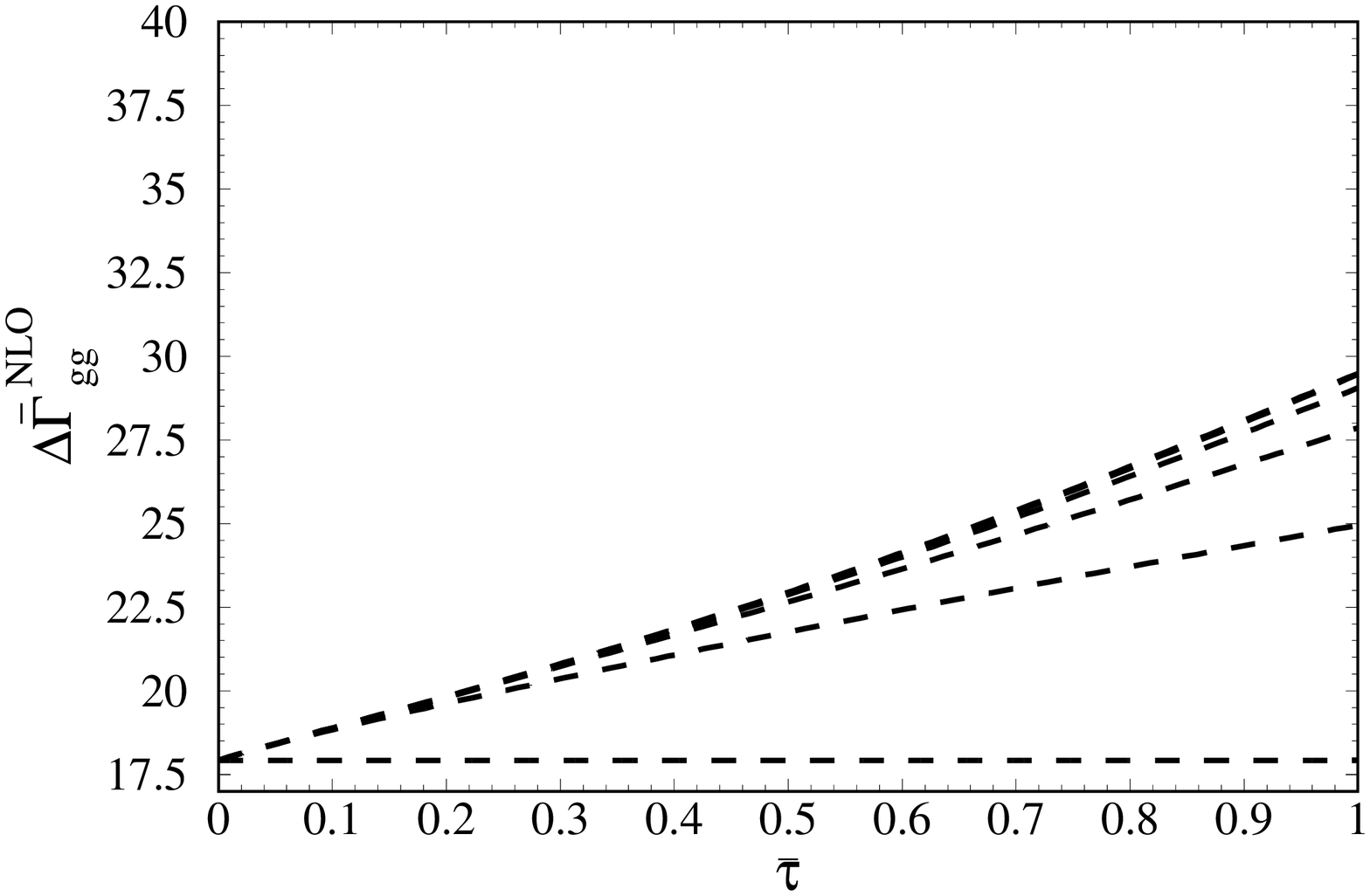}
      &
      \includegraphics[width=0.47\textwidth]{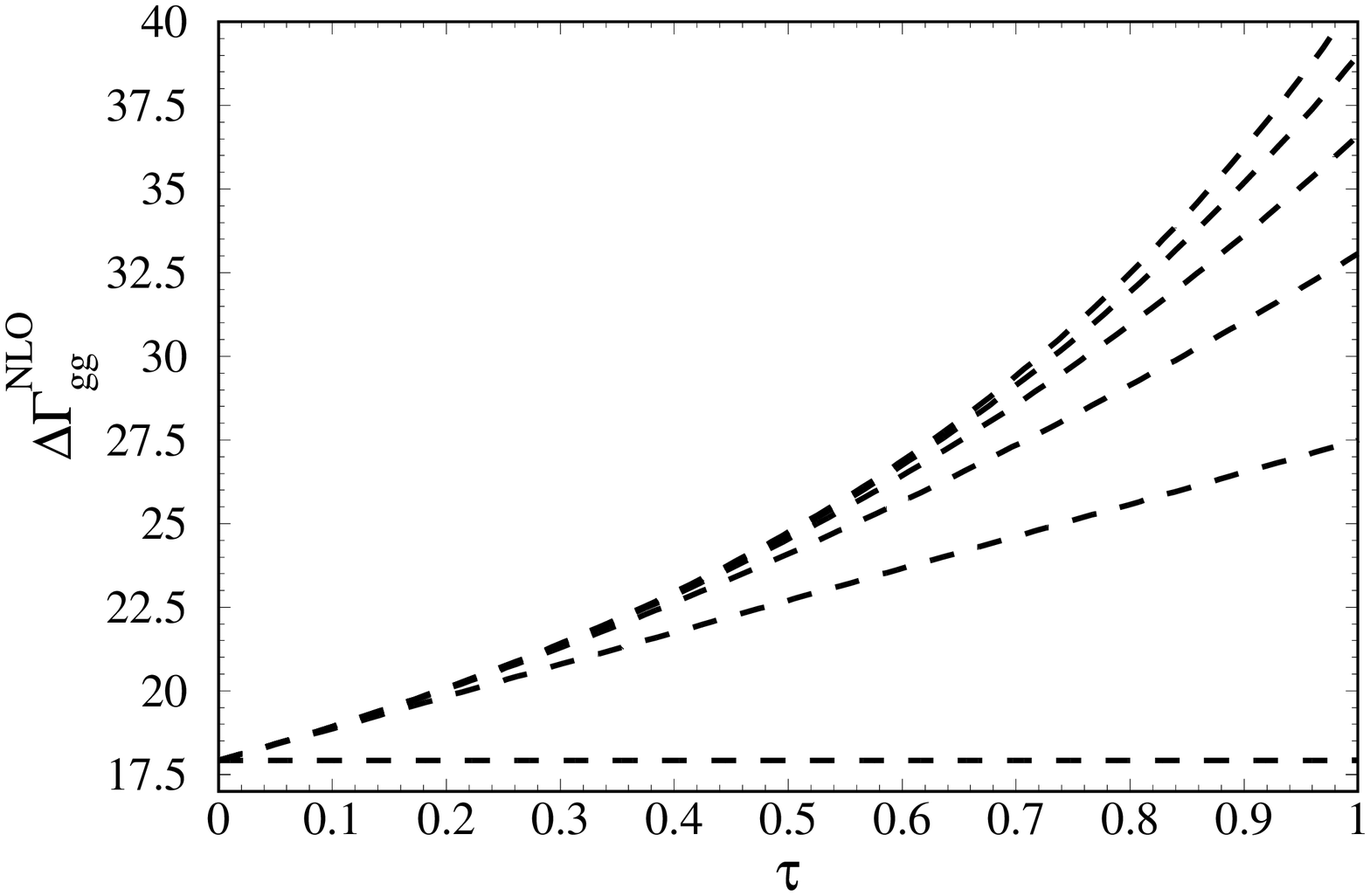}
    \end{tabular}
    \caption{\label{fig::nlo}$\Delta\bar{\Gamma}^{\rm NLO}_{gg}$ (left) and
    $\Delta\Gamma^{\rm NLO}_{gg}$ (right) as a function of 
    $\bar\tau$ and $\tau$, respectively, where
    successively higher orders are included.
    For the renormalization scale $\mu^2=M_H^2$ has been chosen.
    }
  \end{center}
\end{figure}

In Fig.~\ref{fig::lo} $\Gamma^{\rm LO}_{gg}$
is shown as a function of $\tau$ (solid line)
and compared to the approximations 
where successively higher orders are included (dashed lines).
The analog plots for the NLO results
$\Delta\bar{\Gamma}^{\rm NLO}_{gg}$ and $\Delta\Gamma^{\rm NLO}_{gg}$
are shown in Fig.~\ref{fig::nlo} where $\mu^2=M_H^2$ has been chosen.
From the behaviour of the approximations we can deduce that
in the region, where the curves including different orders in the
$M_t$-expansion are on top
of each other, they also coincide with the exact result. Having this in
mind we conclude for the NLO corrections that in case of the pole mass
the approximations including 
the $\tau^5$ ($\tau^2$) terms coincide with the exact
result up to $\tau\approx 0.60\,(0.35)$ 
which corresponds to
$M_H\approx270\,(200)$~GeV. It is worth to mention that the top quark
mass-suppressed corrections become smaller if the $\overline{\rm
  MS}$ mass is used for the parameterization. In this case the 
curves including the $\bar\tau^4$ and $\bar\tau^5$
terms, respectively, are basically on top of each other --- almost up
to $\bar\tau=1$.
Even the curve including corrections up to order $\bar\tau^2$
shows a deviation of less then 6\%
for $\bar\tau=1$ and provides a
perfect approximation up to $\bar\tau=0.5$ (i.e. $M_H\approx250$~GeV). 
Thus, it can be assumed that in the phenomenologically relevant region for
$M_H$ the results including the first three expansion terms in $\tau$
represent an excellent approximation for all practical purposes.
Let us also mention that for $M_H=120$~GeV (corresponding to
$\tau\approx0.1$) the leading term in the $1/M_t$ expansion
approximates the exact result with an accuracy of about 5\% 
--- both for the on-shell and $\overline{\rm MS}$ top quark mass.
These considerations provide a strong motivation to compute the 
first three expansion terms at NNLO.

Very often, in particular in the context of the Higgs boson production
in the gluon fusion process, the complete LO result is factored
out when considering the perturbative expansion. In this case we
define
\begin{eqnarray}
  \Gamma(H\to gg) &=& \Gamma_0 
  \left(\frac{\alpha_s(\mu)}{\pi}\right)^2
  \Gamma^{\rm LO}_{gg}\left[1 
  + \frac{\alpha_s(\mu)}{\pi} \delta\Gamma^{\rm NLO}_{gg}
  + \left(\frac{\alpha_s(\mu)}{\pi}\right)^2\delta\Gamma^{\rm
  NNLO}_{gg} 
  + \ldots
  \right]
  \,.
\end{eqnarray}
The coefficients $\delta\Gamma^{\rm NLO}_{gg}$ --- both in the
$\overline{\rm MS}$ and on-shell scheme --- can easily be obtained
from the results for $\Delta\Gamma^{\rm NLO}_{gg}$. We refrain from
listing them explicitly.

\begin{figure}[t]
  \begin{center}
    \begin{tabular}{cc}
      \includegraphics[width=0.47\textwidth]{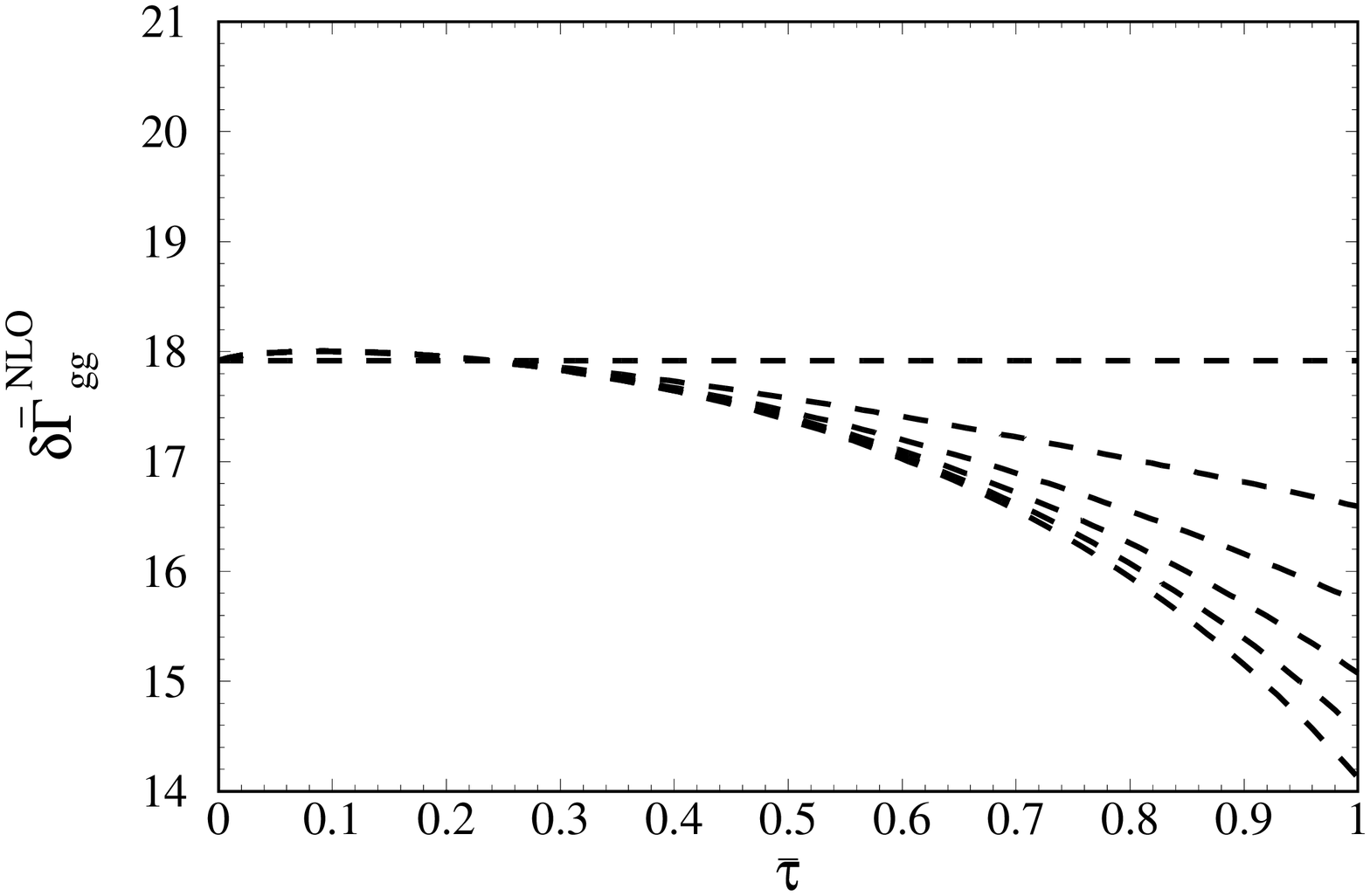}
      &
      \includegraphics[width=0.47\textwidth]{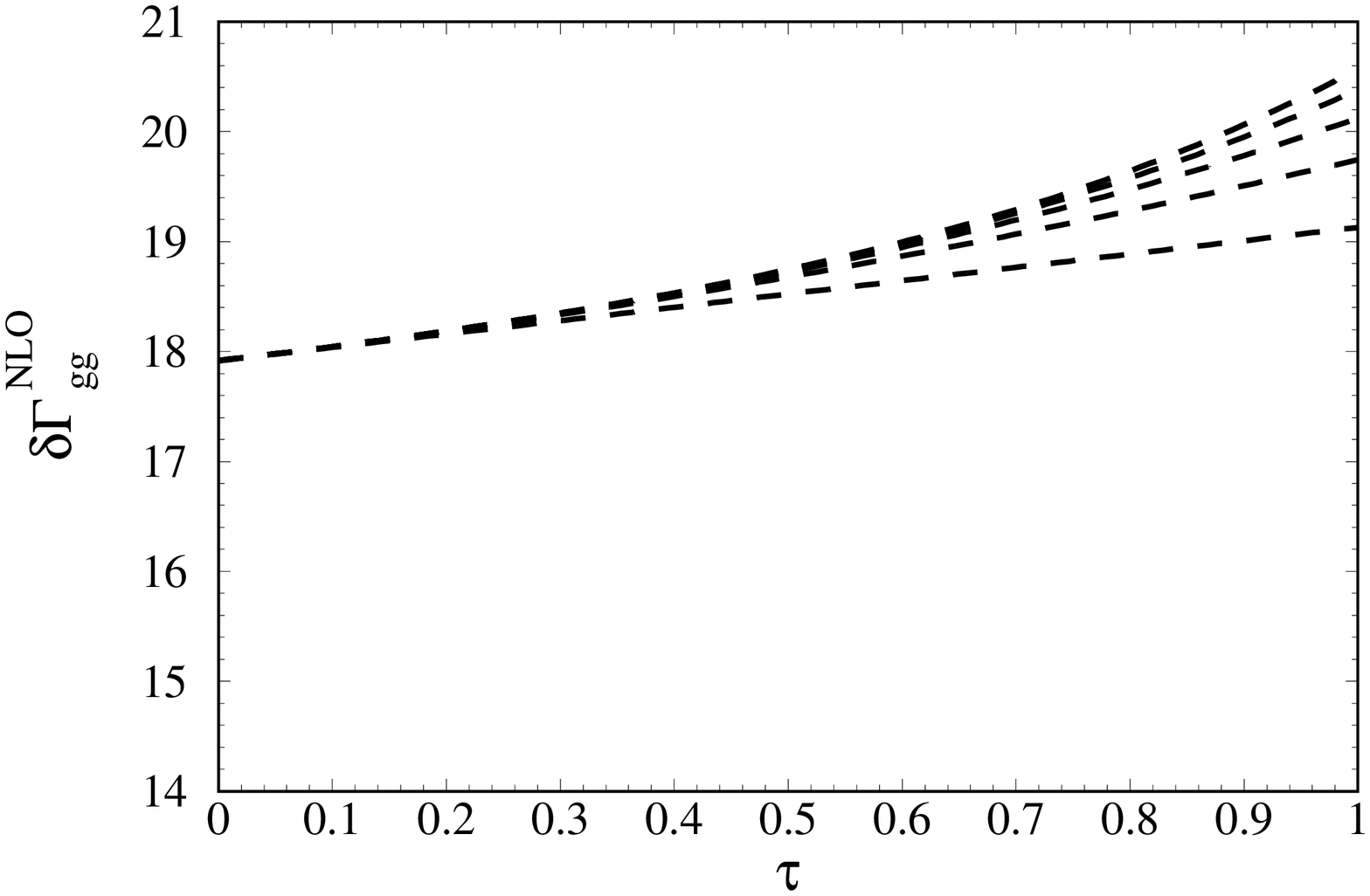}
    \end{tabular}
    \caption{\label{fig::nlo-2}
      $\delta\bar{\Gamma}^{\rm NLO}_{gg}$ (left) and
      $\delta\Gamma^{\rm NLO}_{gg}$ (right) as a function of 
      $\bar\tau$ and $\tau$, respectively, where
    successively higher orders are included.
    For the renormalization scale $\mu^2=M_H^2$ has been chosen.
    }
  \end{center}
\end{figure}

In Fig.~\ref{fig::nlo-2} $\delta\bar\Gamma^{\rm NLO}_{gg}$ and
$\delta\Gamma^{\rm NLO}_{gg}$ are shown as a function of
 $\bar\tau$ and $\tau$, respectively,
where successively higher orders are included. 
One observes a dramatic
improvement in the convergence of the $1/M_t$ expansion for
$\tau\lsim0.3$ where already the leading $M_t$-term is practically
indistinguishable from the exact result.

%- }}}
%- {{{ NNLO:

\subsection{$H\to gg$ to NNLO}

Applying the methods described in Section~\ref{sec::tecdet} to the
five-loop diagrams shown in Fig.~\ref{fig::diags} leads to the NNLO
corrections to the decay rate $\Gamma(H\to gg)$. We were able to
compute three expansion terms in $\tau$ which can be cast in the form
\begin{eqnarray}
  \Delta\bar{\Gamma}^{\rm NNLO}_{gg}
  &=& h_0^{\rm{nnlo}}+h_1^{\rm{nnlo}}\taums+h_2^{\rm{nnlo}}\taums^2
    + \ldots
  \,,
  \label{eq::gamnnlo}
\end{eqnarray}
with
\begin{eqnarray}
  h_0^{\rm nnlo}&=&
  \frac{149533}{288} 
  - \frac{121}{16}\pi^2 
  - \frac{495}{8}\zeta(3)
  + \frac{3301}{16}\LH + \frac{363}{16}\LH^2 
  + \frac{19}{8}\Lt
  + n_l\left(-\frac{4157}{72} 
  + \frac{11}{12}\pi^2 
  \right.\nonumber\\&&\left.\mbox{}
  + \frac{5}{4}\zeta(3)
  - \frac{95}{4}\LH - \frac{11}{4}\LH^2 + \frac{2}{3}\Lt
  \right) 
%  \nonumber\\&&\mbox{}
  + n_l^2\left(\frac{127}{108} - \frac{\pi^2}{36}
  + \frac{7}{12}\LH + \frac{\LH^2}{12} 
  \right) 
  \,,\nonumber\\
  h_1^{\rm nnlo}&=&
  \frac{104358341}{1555200}
  - \frac{847}{240}\pi^2
  + \frac{7560817}{69120}\zeta(3) 
  + \LH\left(\frac{203257}{2160} - \frac{77}{15}\Lt\right)
  + \frac{847}{80}\LH^2
  \nonumber\\&&\mbox{}
  - \frac{24751}{1080}\Lt - \frac{77}{180}\Lt^2 
  + n_l\left[-\frac{9124273}{388800}
  + \frac{77}{180}\pi^2 + \frac{7}{12}\zeta(3)
  + \LH\left(-\frac{67717}{6480} 
  \right.\right.\nonumber\\&&\left.\left.\mbox{}
  + \frac{14}{45}\Lt\right)
  - \frac{77}{60}\LH^2
  + \frac{586}{405}\Lt + \frac{7}{90}\Lt^2
  \right]
  + n_l^2\left(\frac{5597}{12960} 
  - \frac{7}{540}\pi^2
  + \frac{29}{120}\LH 
  \right.\nonumber\\&&\left.\mbox{}
  + \frac{7}{180}\LH^2
  \right)
  \,,\nonumber\\
  h_2^{\rm nnlo}&=&
  - \frac{1279790053883}{12192768000}
  - \frac{186703}{100800}\pi^2
  + \frac{39540255113}{232243200}\zeta(3)
  + \LH\left(\frac{9158957}{189000}  
  \right.\nonumber\\&&\left.\mbox{}
  - \frac{16973}{3150}\Lt\right) 
  + \frac{186703}{33600}\LH^2
  - \frac{10980293}{453600}\Lt 
  + \frac{20059}{37800}\Lt^2 
  + n_l\left[-\frac{64661429393}{5715360000} 
  \right.\nonumber\\&&\left.\mbox{}
  - \frac{16973}{25200}\LH^2
  + \frac{16973}{75600}\pi^2 
  + \frac{1543}{5040}\zeta(3)
  + \LH\left(-\frac{10306537}{1944000} + \frac{1543}{4725}\Lt\right) 
  \right.\nonumber\\&&\left.\mbox{}
  + \frac{8973773}{6804000}\Lt 
  + \frac{1543}{18900}\Lt^2
  \right]
  +  n_l^2\left(\frac{3829289}{19440000}
  - \frac{1543}{226800}\pi^2
  + \frac{89533}{756000}\LH 
  \right.\nonumber\\&&\left.\mbox{}
  + \frac{1543}{75600}\LH^2
  \right) 
  \,,
\label{eq::hgg-nnlo}
\end{eqnarray}
where the $\overline{\rm MS}$ top quark mass has
been used. Transforming the result to the on-shell scheme and
evaluating it in numerical form leads to
\begin{eqnarray}
  \Delta\Gamma^{\rm NNLO}_{gg}
  &=&
    156.808 + 102.146 L_H + 11.021 L_H^2  + 5.708 L_T + 
    \nonumber\\&&\mbox{}
    \left(109.365 + 52.662 L_H + 5.143 L_H^2  + 4.645 L_T\right)\tau + 
    \nonumber\\&&\mbox{}
    \left(74.434 + 29.920 L_H + 2.699 L_H^2  + 3.297 L_T\right) \tau^2
    + \ldots
  \,,
  \label{eq::gamnnlo-os}
\end{eqnarray}
with $L_T=\ln(\mu^2/M_t^2)$.
We want to mention that the result for $h_0^{\rm nnlo}$ has already
been obtained in Ref.~\cite{Chetyrkin:1997iv,Steinhauser:2002rq}, 
the remaining terms are new.

Let us for completeness mention that the leading term of the large top
quark mass expansion is also known at NNNLO~\cite{Baikov:2006ch}. It
is given by 
\begin{eqnarray}
  \Delta\Gamma^{\rm NNNLO}_{gg}
  &=&
  467.684 + 1215.302\, L_H + 394.514\, L_H^2 
    + 28.164\, L_H^3
    \nonumber\\&&\mbox{}
    + 21.882\,L_H L_T 
    + 122.441\, L_T + 10.941\, L_T^2   
    + \ldots
  \,.
  \label{eq::gamnnnlo-os}
\end{eqnarray}

\begin{figure}[t]
  \begin{center}
    \begin{tabular}{cc}
      \includegraphics[width=0.47\textwidth]{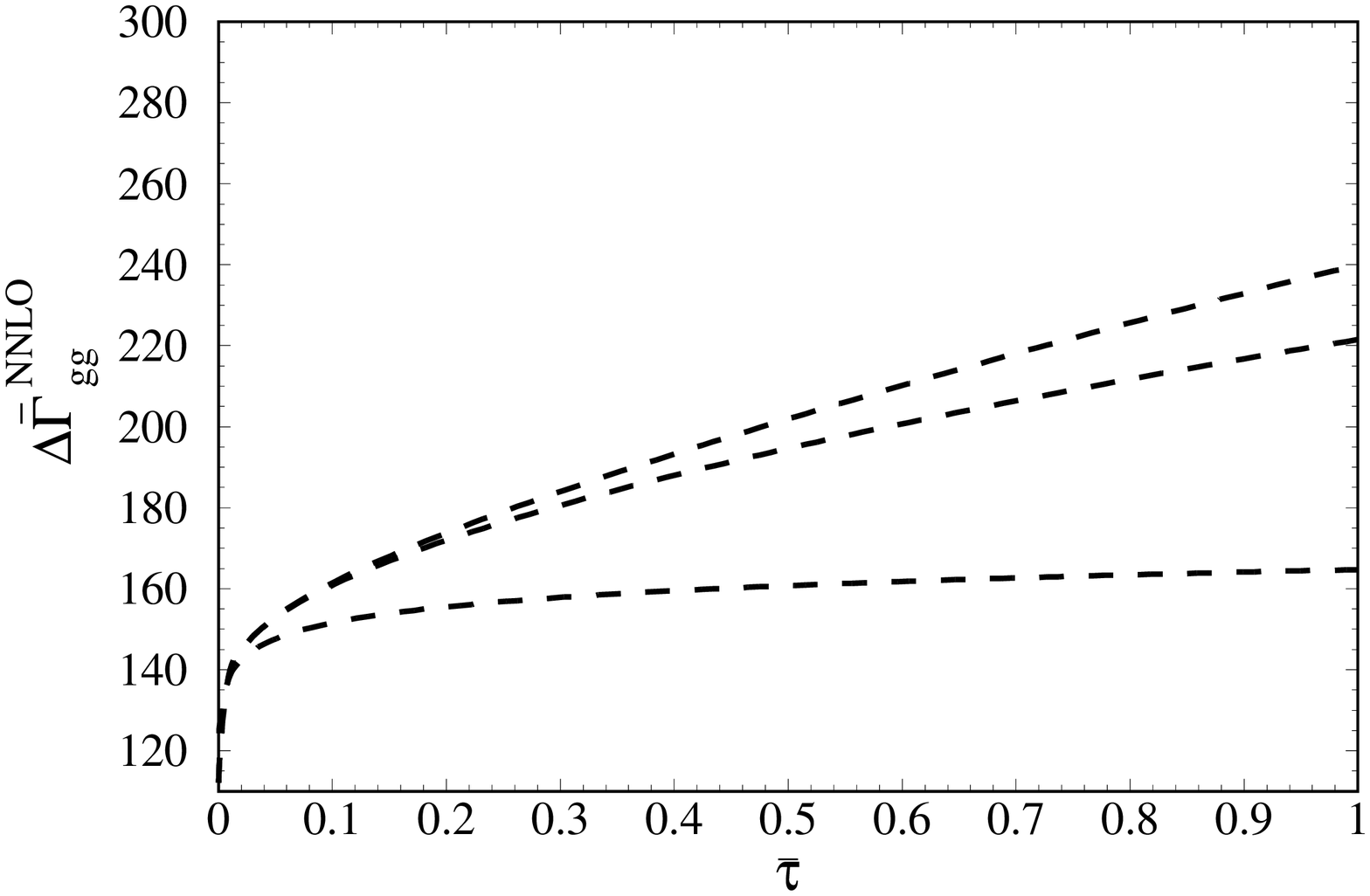}
      &
      \includegraphics[width=0.47\textwidth]{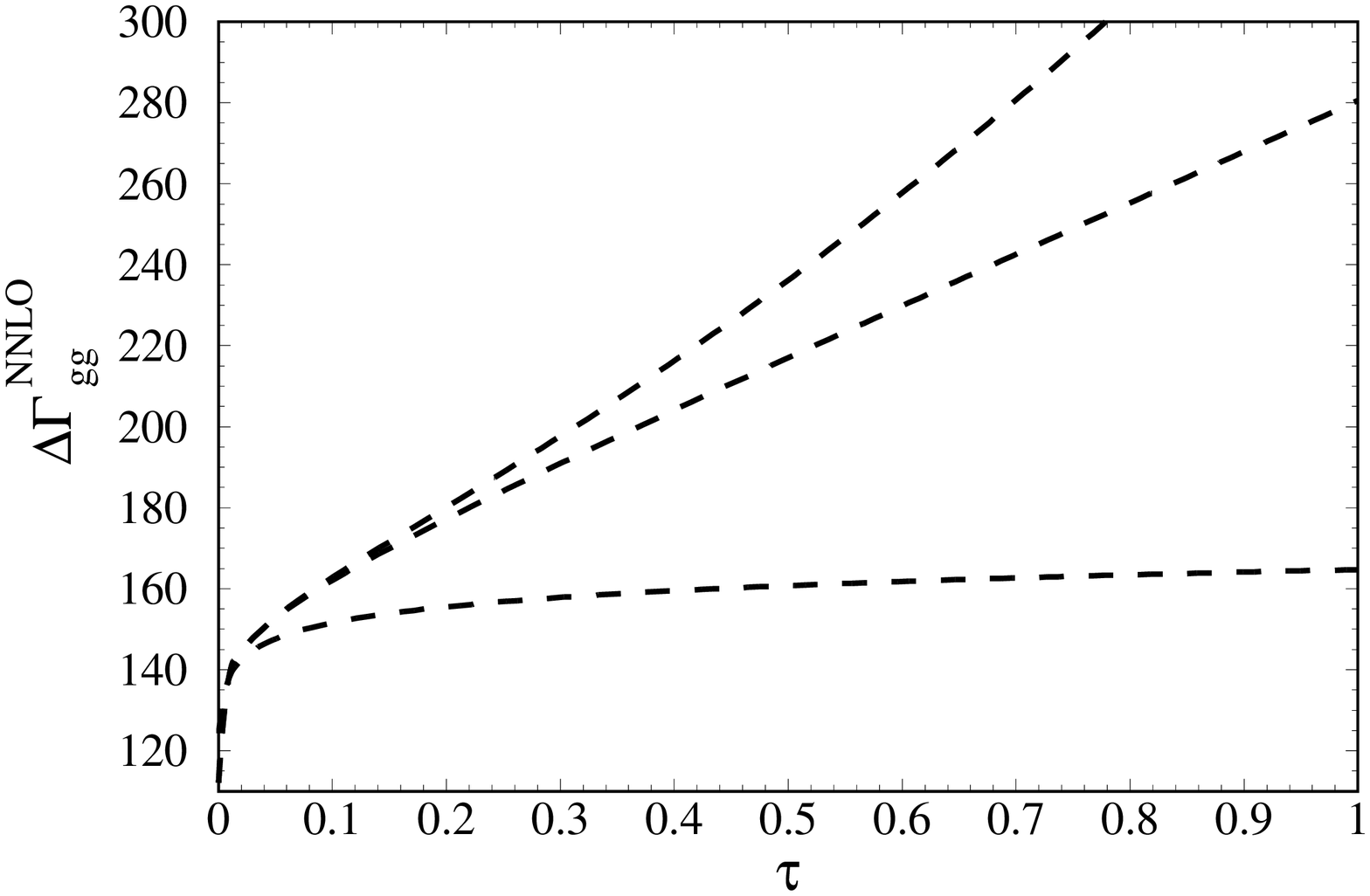}
    \end{tabular}
    \caption{\label{fig::nnlo}
      $\Delta\bar{\Gamma}^{\rm NNLO}_{gg}$ (left) and
      $\Delta\Gamma^{\rm NNLO}_{gg}$ (right) as a function of 
      $\bar\tau$ and $\tau$, respectively, where
    successively higher orders are included.
    For the renormalization scale $\mu^2=M_H^2$ has been chosen.
    }
  \end{center}
\end{figure}

In Fig.~\ref{fig::nnlo} we show the results of
Eqs.~(\ref{eq::gamnnlo}) and~(\ref{eq::gamnnlo-os}).
For small values of $\tau$ one observes a logarithmic divergence 
which even occurs for $\mu=M_H$. In the phenomenologically relevant
range ($\tau\gsim0.1$) the logarithm $\ln (M_H^2/M_t^2)$
is numerically small and we observe
the same pattern as at LO and NLO which is a strong indication that
our result including the $\tau^2$ terms provides an excellent approximation
to the exact result up to $M_H\approx M_t$ ($\tau\approx 0.25$).
For $M_H=120$~GeV ($\tau\approx0.1$) we observe a correction of about
9\% 
originating from the power-suppressed terms. This induces an
effect of approximately 1\%
on the decay rate $\Gamma(H\to gg)$.

\begin{figure}[t]
  \begin{center}
    \begin{tabular}{cc}
      \includegraphics[width=0.47\textwidth]{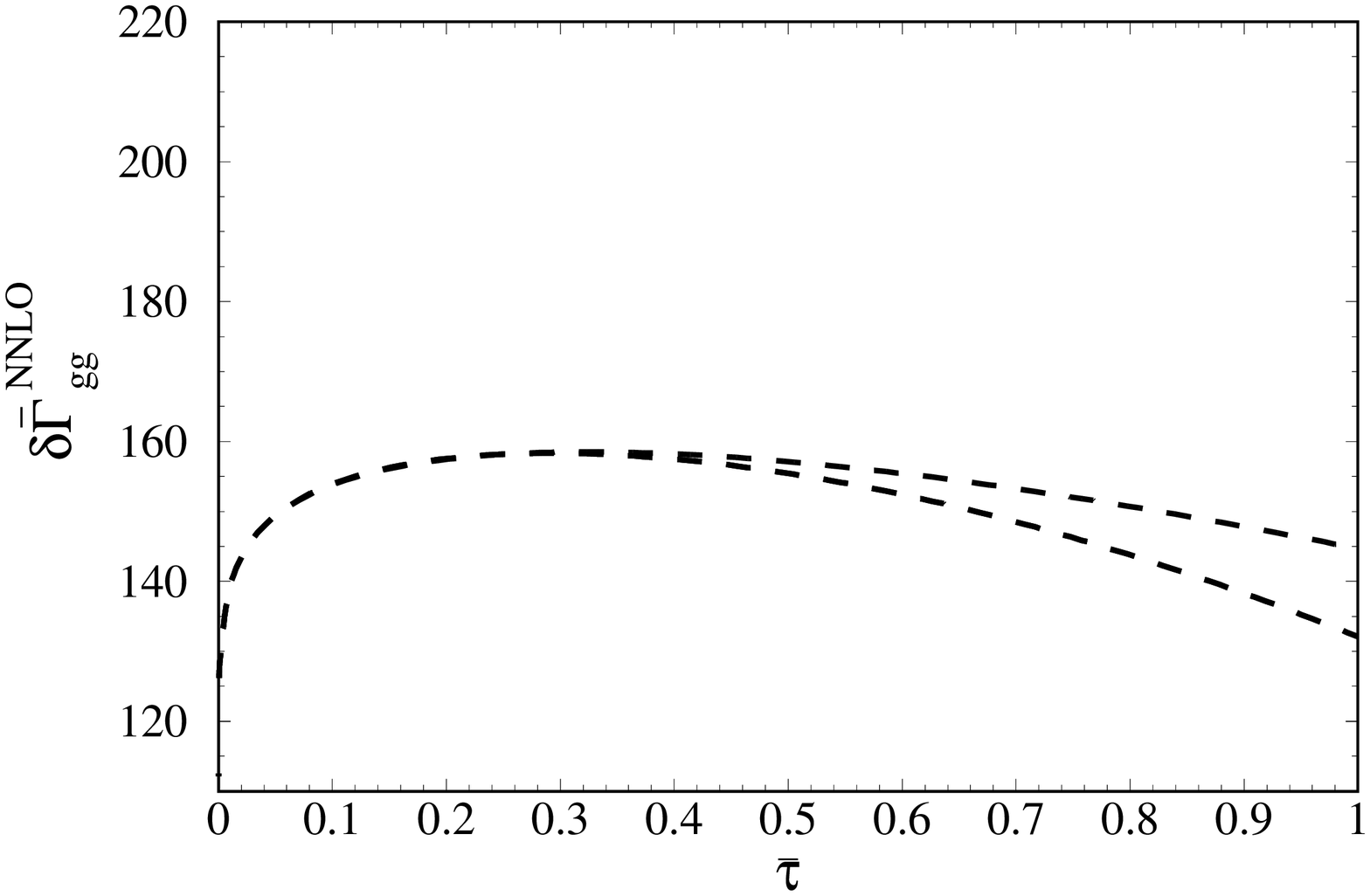}
      &
      \includegraphics[width=0.47\textwidth]{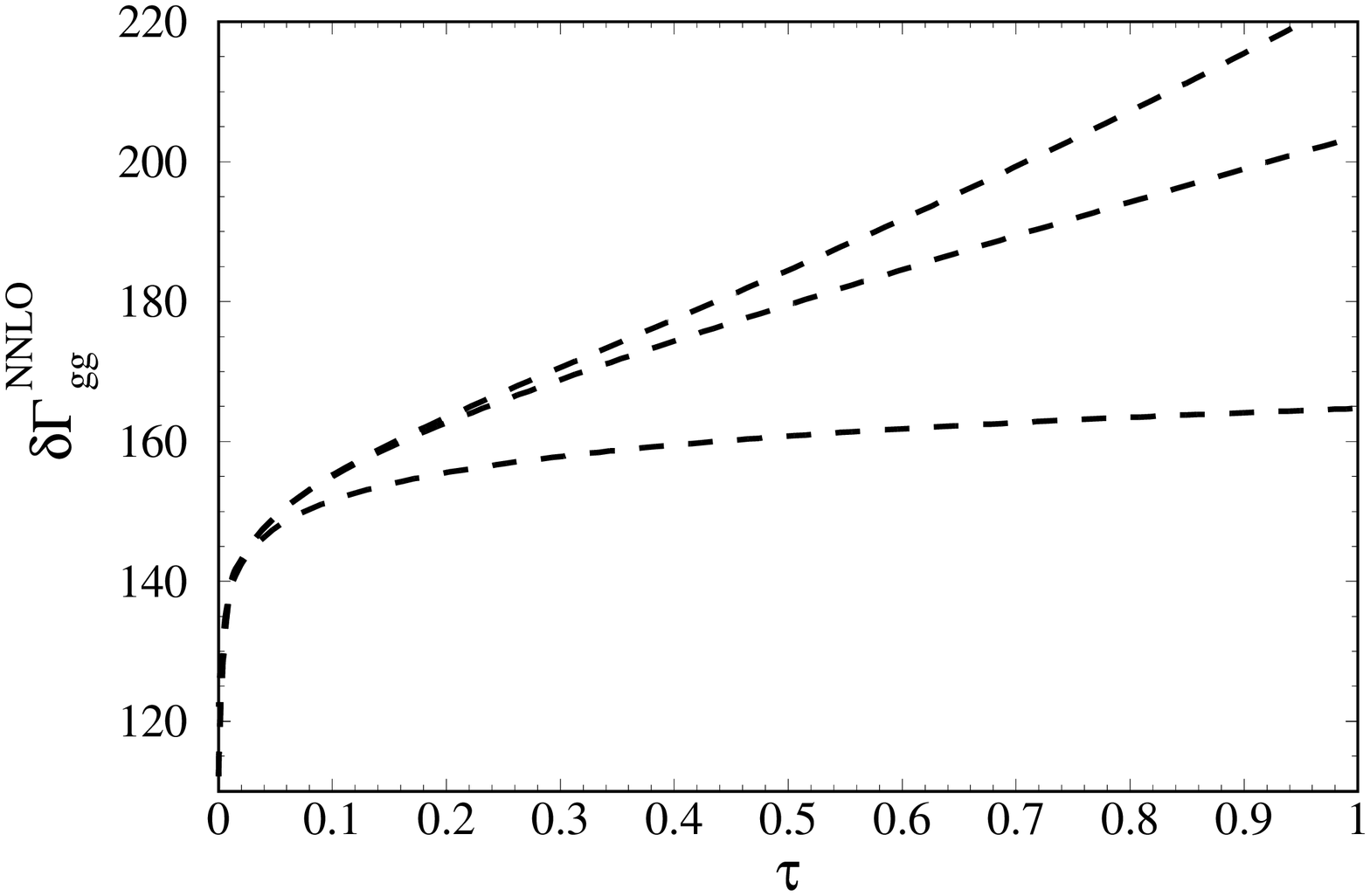}
    \end{tabular}
    \caption{\label{fig::nnlo-fac}
      $\delta\bar{\Gamma}^{\rm NNLO}_{gg}$ (left) and
      $\delta\Gamma^{\rm NNLO}_{gg}$ (right) as a function of 
      $\bar\tau$ and $\tau$, respectively, where
    successively higher orders are included.
    For the renormalization scale $\mu^2=M_H^2$ has been chosen.
    }
  \end{center}
\end{figure}

Fig.~\ref{fig::nnlo-fac} shows the result for
$\delta\bar\Gamma^{\rm NNLO}$ and $\delta\Gamma^{\rm NNLO}$. Similarly to the
NLO result one observes a reduction of the size of the mass
corrections. This is particularly pronounced in the case of the
$\overline{\rm MS}$ scheme where the curves involving terms up to
order $\bar\tau^1$ and $\bar\tau^2$ lie on top of each other and coincide with 
the leading term up to $\bar\tau\approx0.4$ ($M_H\approx220$~GeV).
These considerations provide a strong motivation to consider the total
decay rate where the complete LO result is extracted. Furthermore,
due to the similarity to the production of Higgs bosons in the gluon
fusion channel it can be expected that there the mass corrections are
also at the few percent level --- below the current uncertainties from
the parton distribution
functions~\cite{Harlander:2002wh,Anastasiou:2002yz,Ravindran:2003um,Moch:2005ky}.

Let us finally mention that the numerical effect of the 
power-suppressed terms computed in this paper are comparable to the
leading $M_t$-term at NNNLO (cf. Eq.~(\ref{eq::gamnnnlo-os})).

\begin{figure}[t]
  \begin{center}
    \begin{tabular}{c}
      \includegraphics[width=0.95\textwidth]{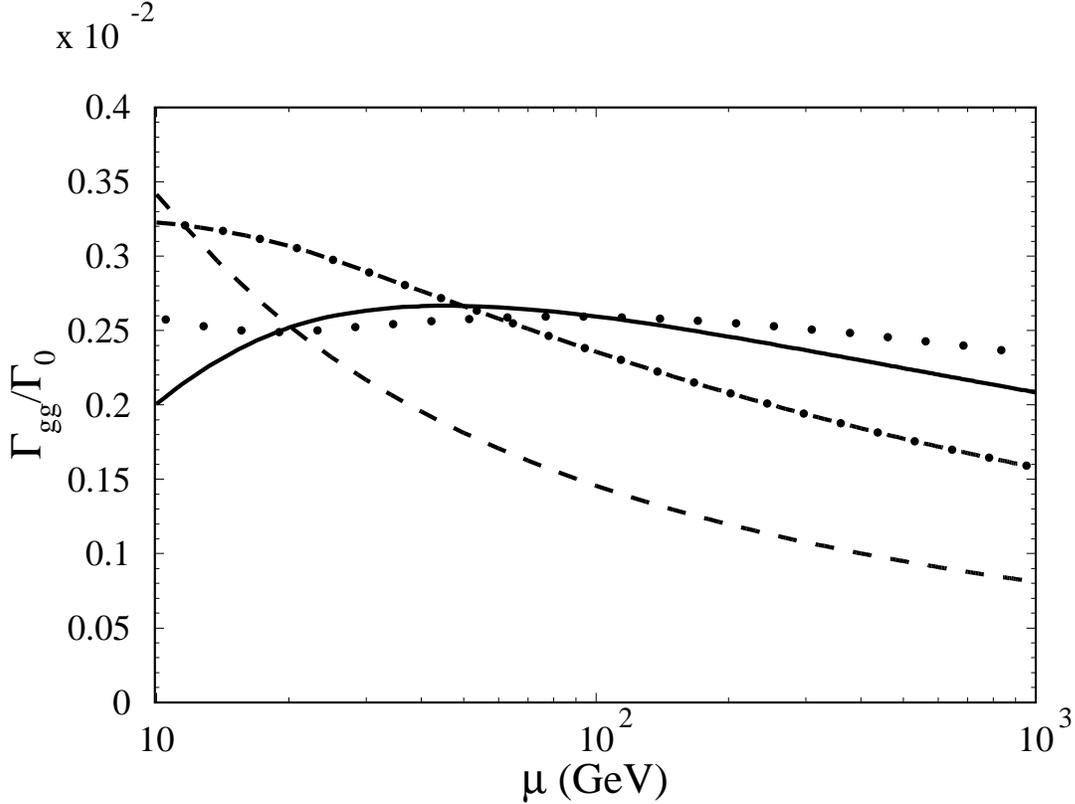}
    \end{tabular}
    \caption{\label{fig::mu}
      $\Gamma_{gg}/\Gamma_0\equiv
      \Gamma(H\to gg)/\Gamma_0$ as a function of $\mu$
      with $\Gamma_0=G_F M_H^3/(36\pi\sqrt{2})$. The LO, NLO and NNLO
      results are shown as dashed, dash-dotted and solid lines.
      The dotted curve also includes the leading top quark mass
      dependent term at NNNLO~\cite{Baikov:2006ch}.
    }
  \end{center}
\end{figure}

At this point it is interesting to consider the dependence of
$\Gamma(H\to gg)$ on the renormalization scale $\mu$. In
Fig.~\ref{fig::mu} the LO (dashed), NLO (dash-dotted),
NNLO (solid) and NNNLO (dotted) result is shown for $\mu$
between 10~GeV and 1~TeV where for the LO curve the exact result has
been used. The NLO and NNLO curve include terms of order $\tau^5$ and
$\tau^2$, respectively. As input for our numerical studies we use
$\alpha_s(M_Z)=0.118$, $M_Z=91.1876$~GeV~\cite{PDG:2007},
$M_t=170.9$~GeV~\cite{unknown:2007bx} and $M_H=120$~GeV.
The renormalization group equations are solved with the help of the
program {\tt RunDec}~\cite{Chetyrkin:2000yt}.

It is interesting to note that around $\mu=40$~GeV
the NNLO curve shows a local maximum where the decay rate is
independent of $\mu$. 
For $\mu=50$~GeV the NNLO corrections vanish (crossing between
dash-dotted and solid line) which hints for a good convergence of
perturbation theory.
This is further supported by the smallness of the NNNLO corrections
(dotted) which show a very flat $\mu$-dependence over the whole 
range considered in Fig.~\ref{fig::mu}.

%- }}}

%- }}}
%- {{{ Conclusions:

\section{\label{sec::concl}Conclusions}

In this paper the NNLO corrections to the gluonic decay width of the
Higgs boson in the intermediate mass range is considered. With the
help of an automated asymptotic expansion three terms in the
large-$M_t$-limit have been obtained. It is demonstrated that our
result is equivalent to an exact calculation at least up to $M_H=M_t$.
For $M_H=120$~GeV the NNLO term which changes by about 9\% 
due to the new corrections induces a 1\%
correction to the total gluonic decay rate. 
The situation is different if the complete top quark mass dependence of
the leading order result is factored out. In this case the power-suppressed
corrections become smaller. This is in particular true for top quark masses in the
$\overline{\rm MS}$ scheme where for Higgs boson masses up to $220$~GeV the full result is given
by the leading term in the large $M_t$-expansion. This observation has possible implications
for the Higgs boson production via gluon fusion where only the leading order term
in $M_H/M_t$ is available. 
In case the expansion for $M_H\ll M_t$ shows a similar
behaviour the uncertainty induced by the unknown 
power-suppressed terms is negligible.

%- }}}
%- {{{ Acknowledgements:

\bigskip
\noindent
{\large\bf Acknowledgements}\\
We would like the thank Robert Harlander for useful comments
and carefully reading the manuscript.
This work was supported by the BMBF through 05HT6VKA
and by the DFG through SFB/TR~9.

%- }}}
%- {{{ Bibliography:

%- }}}

\end{document}